# Toward Effective PBFT Consensus Service under Software Aging in Dynamic Scenarios

Yujing Cai, Yukun Meng, Weimeng Wang, Xuanming Zhang, Xiaolin Chang

**ABSTRACT** The increasing application and deployment of blockchain in various services necessitates the assurance of the effectiveness of PBFT (Practical Byzantine Fault Tolerance) consensus service. The paper explores how to reduce the consensus processing time and maintenance cost of PBFT consensus service under non-Byzantine errors. We first propose a PBFT system, consisting of three subsystems, one active-node subsystem, one standby-node subsystem and a repair subsystem. All the active nodes participate the consensus and all standby nodes aim for fault-tolerance. Each non-Byzantine-failed nodes become standby nodes after completing its repairing in the repair subsystem. The nodes migrate between the active-node and standby-node subsystems in order to support the continuity of the PBFT consensus service while reducing maintenance cost. Then, we develop a Markov-chain-based analytical model for capturing the behaviors of the system and also derive the formulas for calculating the metrics, including consensus processing time, PBFT service availability, the mean number of nodes in each subsystem. Finally, we design a Multi-Objective Evolutionary Algorithm-based method for minimizing both the PBFT serice response time and the PBFT system maintenance cost. We also conduct experiments for evaluation.

## I. INTRODUCTION

Consortium Blockchain, referred to as a distributed ledger comprised of a series of blocks, has been regarded as a transformative technology across a multitude of industries like finance [1], healthcare [2]-[3], energy [4]-[6], logistics [7]-[8], and security defense [9]-[10]. The consensus mechanisms, as the foundation of the blockchain, aim to assist participating nodes to attach blocks to the blockchain in the pre-defined order by reaching an agreement. Two types of errors can occur to consensus nodes, Byzantine and non-Byzantine errors [11][12]. Byzantine errors (Byzantine Fault) mean that consensus nodes can forge or tamper with the information and respond maliciously. Non-Byzantine errors (Crash Fault) mean that a consensus node fails to make a response, maybe due to software aging [18].

The Practical Byzantine Fault Tolerance (PBFT) consensus mechanism is a foundational protocol in the fields of distributed systems [13]. It is designed to maintain reliability and security when up to one-third of the consensus nodes are compromised or behaving arbitrarily [14]. Therefore, PBFT is particularly suitable for the environments where high security and fault tolerance are both of significant importance. It has been applied in blockchain platforms like Hyperledger Fabric [15] and it is a viable option for enterprise-level deployments where there is critical demand for performance and scalability.

The past years have witnessed significant researches in the PBFT development to effectively improve the PBFT's performance and reliability [16][17]. But there is a lack of system flexibility in most existing PBFT-related researches. That is, the consensus nodes are not allowed freely join or exit the PBFT network. It is noticed that any software suffers from software aging after a long and continuous running. Therefore, non-Byzantine errors cannot be avoided.

The paper explore an effective PBFT system under non-Bazantine errors. By effective, we mean the reduction of both the consensus processing time and maintenance cost of the system. Fig.1 illustrates the system proposed in this paper, consisting of three subsystems, one active-node subsystem (denoted hot pool, HP), one standy-node subsystem (denoted warm pool, WP) and a repair subsystem (denoted repair pool, RP). All the active nodes participate the consensus. If an active node crashes, it enters into the repair subsystem and then enters into the WP after the completion of its repairing. In addition, nodes migrate between HP and WP at a certain rate in order to reduce maintenance cost without sacrificing the consensus performance in a dynamic environment. We apply the techniques of analytical modeling [18] and evolutionary algorithm [19][33] to minimize the consensus processing time and maintenance cost of a PBFT system under non-Bazantine errors. To the best of our konwledge, we are the first for studying such a system and improving the system performance with the cost as less as possible.

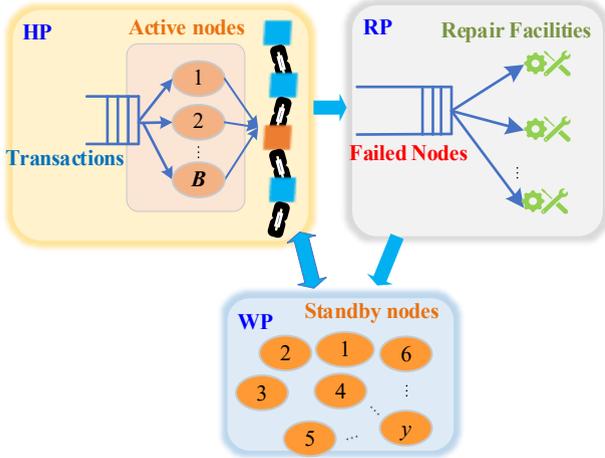

Fig. 1 Steps of the event process

The main contributions of this paper are summarized as follows:
- We propose a PBFT system (detailed in Section III.A), including three subsystems which collaborate to complete the consensus in time without causing much maintenance cost in the dynamic environment. This maintenance cost include the node cost in HP and WP subsystems, the repairing cost and the migration cost between the HP and WP subsystems.
- We develop a novel Markov-chain-based analytical model and the metric formulas for the quantitative analysis of the PBFT system performance and availability. We detail the model and the formulas for calculating the metrics. This analysis is crucial for understanding the impact of different parameters and for making informed decisions to enhance the efficiency and resilience of PBFT implementations.
- We apply the deveopled metric formulas to establish the optimization problem for reducing both the PBFT service response time and the PBFT system maintenance cost. The PBFT system maintenance cost is composed of the node working cost, node migration cost and node repairing cost.
- We design a *N*ondominated *S*orting *G*enetic *A*lgorithm II (NSGA-II)-based method for the optimization problem. NSGA-II is one of the most prominent multiobjective EA (MOEA) with many successful applications in various field [33]. The method is detailed in Section IV.B.

This paper uses transaction and block interchangeably. The rest of the paper is organized as follows. In Section II, we provide an overview of related work. Section III introduces the proposed model and the metric formulas. Section IV discusses the experimental results. Finally, we conclude this paper and possible directions for future work in Section V.

## II. RELATED WORK

This section focuses on the existing research methods for the performance evaluation of blockchain systems. The existing mainstream blockchain performance evaluation methods fall into two major categories: empirical analysis and analytical modeling [20]. At present, the empirical evaluation methods of blockchain include benchmarking [21], monitoring [22], experimental analysis [23] and so on. Our paper utilizes the second type approach, analytical modeling, to study PBFT consensus. Therefore, the following focuses on the analytical-modeling-based works.

Pongnumkul et al. [24] developed stochastic models for evaluating Hyperledger Fabric and Ethereum. Their results indicated that Hyperledger Fabric consistently outperforms Ethereum across all evaluation metrics, including execution time, latency, and throughput. Sukhwani et al. [25] modeled the PBFT consensus process using Stochastic Reward Nets to compute the mean time required to complete consensus for networks with a maximum of 100 peers. Jiang et al. [15] considered two timeout constraints and designed a hierarchical model for Hyperledger Fabric v1.4 transaction process from the client's transaction submitting to the completion of validating/committing transaction. They also designed the formulas of performance measures, including platform throughput, transaction rejection probability and mean transaction response delay. Ma et al. [26] exploited a two-dimensional Markov process to describe the voting process. Lorünser et al. [27] developed the performance model for PBFT in order to investigate the impact of unreliable channels and the use of different transport protocols over them.

Misic et al. [16][28] proposed the extended PBFT consensus mechanisms for the Internet of Things systems and also developed the corresponding stochastic models for analysis. Qushtom et al. [29] proposed a consensus mechanism by combining PoS and PBFT, which aims to handle dishonest nodes, both individual validators and leaders, while keeping high performance. They also developed a semi-Markov-process-based analytical model for performance evaluation.

All the above analytical-modeling-based works did not consider non-Byzantine errors. Recently, the authors in [17][30] studied the PBFT performance with repairing facilities. **There are at least the following two differences from our work**.
1) The PBFT system is different. We propose to use standby nodes to support the continuity of PBFT services in the dynamic environment. The existing works proposed to use the repaired nodes for the continuity. Therefore, when the system is in the urgent need of consensus nodes, only the repairing rate can be adjusted to make up for the lack of nodes. But in our work, besides repairing rate, we could also improve the migrating rate of the standby nodes, which have low maintenance cost.
2) We explore how to apply the quantitative analysis to optimize the system performance and cost.

## III. SYSTEM DESCRIPTION AND MODEL

In this section, we first present the system under consideration in this paper. Then its stochastic modeling approach is presented. Finally, the formulas for calculating consensus processing delay and PBFT consensus availability. Table 1 delineates the definitions of the variables to be used.

TABLE 1 Definition of Variables

| Var. | Definition | Default Value |
|---|---|---|
| $h$ | The number of nodes in HP | -- |
| $w$ | The number of nodes in WP | -- |
| $r$ | The number of nodes in RP | -- |
| $q$ | The number of transactions in the queue of HP subsystem | -- |
| $f$ | The maximum number of Byzantine nodes | 3 |
| $B$ | $B=3f+1$ | 10 |
| $N$ | The sum of the number of nodes in HP, WP, and RP, $N=h+w+r \geq 3f+1$ | 15 |
| $K$ | The maximum number of transactions that the HP subsystem can accommodate. | 20 |
| $\lambda$ | The arrival rate of transactions at the HP. | 4 transactions/ms |
| $\mu_h$ | The processing rate of a consensus transaction in the HP. | 5 transactions/ms |
| $\xi$ | The failure rate of a HP node. | 0.5 nodes/ms |
| $c_h, c_w, c_r$ | The maintenance cost a HP/WP/RP node, respectively. | 5,3,2 |
| $c_{hw}, c_{wh}$ | The migration cost from HP to WP, and from WP to HP, respectively | 1, 1.5 |
| $\mu_r$ | The rate of repairing a failed node in the RP. | 10 nodes/ms |
| $\beta_h$ | The rate at which the node moves from HP to WP. | 0.2 nodes/ms |
| $\beta_w$ | The rate at which the node moves from WP to HP. | 8 nodes/ms |
| $\bar{h}, \bar{w}, \bar{r}$ | The mean value of $h, w, r$, respectively. | -- |
| $T_{res}, \bar{T}_{res}$ | Response time and th of the consensus time of a transaction | -- |

### A. System description

Fig.2 illustrates the framework proposed in this paper for the PBFT system, composed of HP subsystem, RP subsystem and WP subsystem. The nodes in HP aim for the PBFT consensus for each arriving transaction request. under the First-come, First-served (FCFS) policy. Each newly-arriving transaction requests will wait in the queue of the HP if all the HP nodes are busy for a consensus. In a PBFT system, there exist Byzantine nodes, which affect the handling of the arriving transactions. Assume that there are at most $f$ Byzantine nodes. According to the PBFT protocol implementation, there are at most $3f+1$ nodes in the HP such that the transaction consensus can be completed correctly. That is, usually $h=3f+1$.

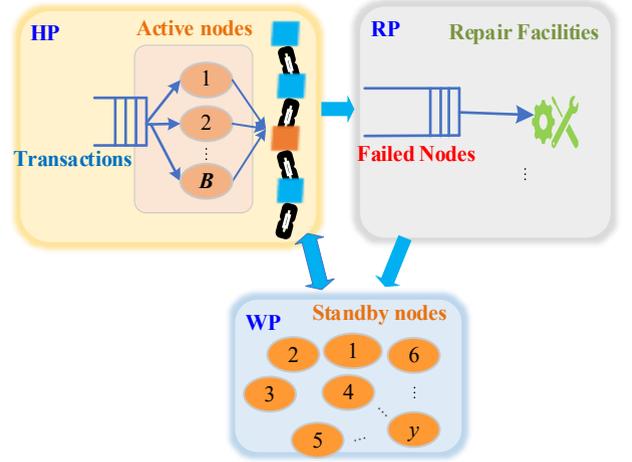

Fig. 2 Illustration of the PBFT system under consideration

However, there exist non-Byzantine errors, such as software aging, which can lead to node crash. Thus, there should be more than $3f+1$ HP nodes in order to keep the continuity of the consensus. We assume that the system has a total of $N$ nodes, which is the sum of the number of nodes in HP, WP, and RP subsystems. Thus, the value of $h$ ranges from $3f+1$ to $N$, depending on the node-failure rate $\xi$.

When $\xi$ is very small, some nodes can be inactive and then save the maintenance cost. Such inactive nodes consist of the WP. When $\xi$ increases, some nodes can move from the WP subsystem to the HP subsystem. Therefore, in the dynamic scenario, nodes can move between the HP and the WP dynamically to assure the PBFT consensus performance. $\beta_h$ is the rate at which the node moves from the HP to the WP, and $\beta_w$ is the rate at which the node moves from the WP to the HP to maintain the PBFT performance. These two variables should be set carefully in order to maintain enough nodes in HP to make consensus.

When non-Byzantine error is detected to occur to a HP node, this node will enter into the RP for fixing/repairing. We define $\mu_r$ as the repair rate of the repair facility in the RP. When a failing node completes its repairing, it enters in the WP subsystem.

In this paper, we assume that transaction arrivals follow a Poisson process, while the inter-event intervals for other events within the system are independent and follow an exponential distribution. We aim to make the quantitative analysis to understand how to set the values of the variables in order to make a tradeoff between PBFT effectiveness and maintenance cost.

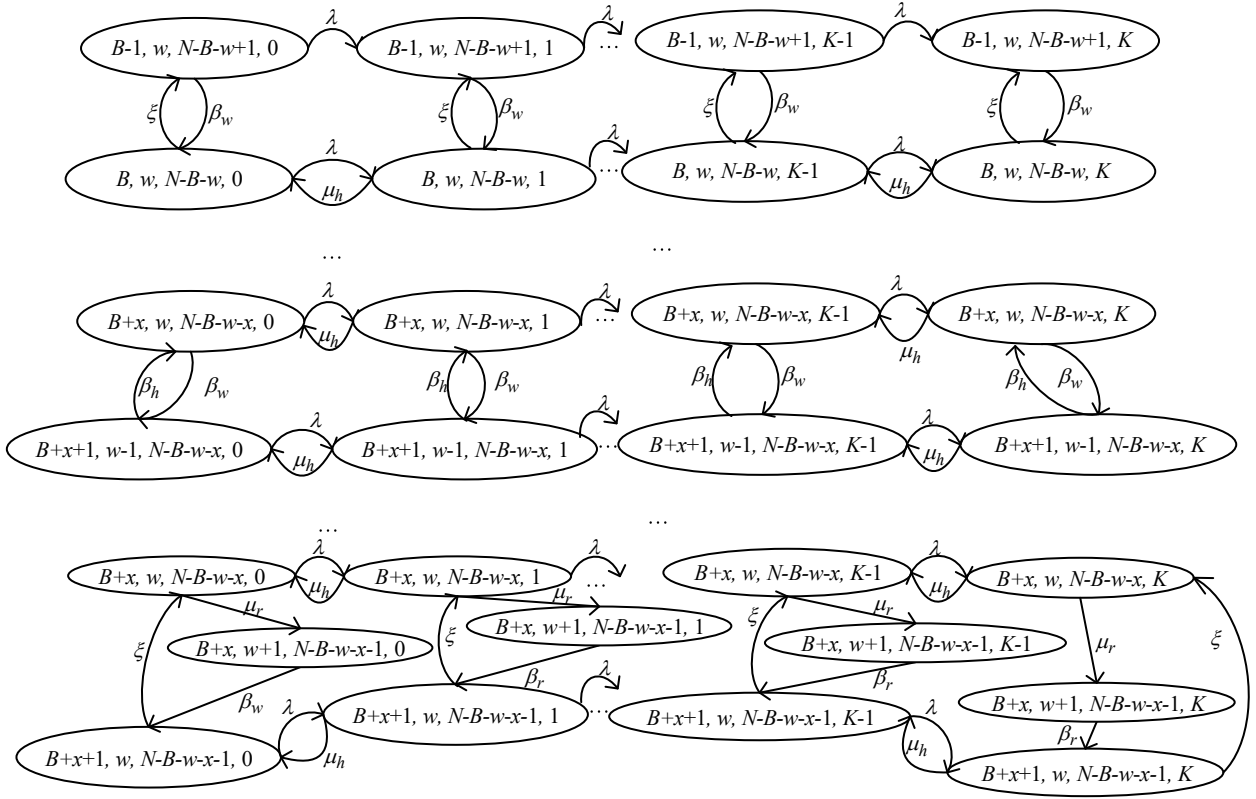

Fig. 3 Illustration of states transitions

## B. Stochastic Model

In light of the aforementioned definitions and assumptions, we now proceed to present the model. Each state is represented by a 4-tuple index $(h, w, r, q)$, as illustrated in Fig.2. Here, the variables $h, w, r$ denote the number of nodes in the HP, WP and RP subsystems, respectively. Their setting are all in $[0, N]$ and $h + w + r = N$. $q$ denotes the number of requests in the HP subsystem.

Now we describe the state transition rules of the following five types of events: 1) a new transaction request arrival, 2) the completion of a consensus, 3) the failure of a HP node, 4) the completion of a RP node, 5) HP node migration, and 6) WP node migration. Fig.3 illustrates some state transitions. We detail the transition rules under each type event in the following.

**Event 1.** New request arrival

Upon the arrival of a request, the resulting state transitions can be classified into two cases according to $q$.

**Case 1.1.** $0 \leq q < K$. The waiting queue of the HP subsystem is not full. Then the arriving transaction request can enter the HP subsystem. The state transition is given by Equation (1).

$$(h, w, r, q) \xrightarrow{\lambda} (h, w, r, q+1) \qquad (1)$$

**Case 1.2.** $q = K$. The waiting queue of the HP subsystem is full. Then the arriving transaction request is dropped. The state transition is given by Equation (2).

$$(h, w, r, K) \xrightarrow{\lambda} (h, w, r, K) \qquad (2)$$

**Event 2.** Consensus completion

When $h \geq 3f + 1$ and $q > 0$, the state transition is given by Equation (3).

$$(h, w, r, q) \xrightarrow{\mu_h} (h, w, r, q-1) \qquad (3)$$

**Event 3.** Failure occurs to a HP node.

The state transition is given by Equation (4) under the condition of h>0

$$(h, w, r, q) \xrightarrow{h \cdot \xi_r} (h-1, w, r+1, q) \qquad (4)$$

**Event 4.** The repairing of a RP node is completed

The state transition is given by Equation (5) under the condition of .

$$(h, w, r, q) \xrightarrow{\mu_r} (h, w+1, r-1, q) \qquad (5)$$

**Event 5. HP node migration**

Only when $h > 3f + 1$, there is HP node migration, given by Equation (6).

$$(h,w,r,q) \xrightarrow{\beta_h} (h-1,w+1,r,q) \quad (6)$$

**Event 6. WP node migration**

The state transition is given by Equation (7).
$$(h,w,r,q) \xrightarrow{\beta_w} (h+1,w-1,r,q) \quad (7)$$

### C. Performance Measures

In accordance with the state transition rules in Section II.B, we derive the state transition rate matrix $Q$ and balance equations, described in Eq. (8).
$$\Pi Q = 0 \quad (8)$$

Here, $\Pi = \{\pi_{(h,w,r,q)} | 0 \leq h,w,r \leq N, 0 \leq q \leq K\}$. The balance equations are employed to compute all $\pi_{(h,w,r,q)}$ with the normalization equation $\sum_{h=0}^{N}\sum_{w=0}^{N}\sum_{r=0}^{N}\sum_{q=0}^{K} \pi_{(h,w,r,q)} = 1$. Note that all $\pi_{(h,w,r,q)}$ with $h+w+r \neq N$ are set to zero because they are meaningless. As these equations cannot be solved in closed form, a numerical solution must be applied. With the obtained $\pi_{(h,w,r,q)}$, we are able to compute the following measures:

Mean number of requests in the authentication service system:
$$E[Q] = \sum_{h=0}^{N}\sum_{w=0}^{N}\sum_{r=0}^{N}\sum_{q=0}^{K} q \cdot \pi_{(h,w,r,q)} \quad (9)$$

Probability of a consensus request being dropped due to the full queue:
$$P_{drop} = \sum_{h=0}^{N}\sum_{w=0}^{N}\sum_{r=0}^{N}\sum_{q=K} \pi_{(h,w,r,q)} \quad (10)$$

Mean delay for processing a consensus request:
$$\overline{T}_{resp} = E[T_{resp}] = \frac{E[Q]}{\lambda \cdot (1-P_{drop})}. \quad (11)$$

The availability of PBFT consensus service:
$$Avail = \sum_{h=3f+1}^{N}\sum_{w=0}^{N}\sum_{r=0}^{N}\sum_{q=0}^{K} \pi_{(h,w,r,q)} \quad (12)$$

The mean number of nodes in each subsystem:
$$\overline{h} = E[h] = \sum_{h=0}^{N}\sum_{w=0}^{N}\sum_{r=0}^{N}\sum_{q=0}^{K} h \cdot \pi_{(h,w,r,q)} \quad (13)$$

$$\overline{w} = E[w] = \sum_{h=0}^{N}\sum_{w=0}^{N}\sum_{r=0}^{N}\sum_{q=0}^{K} w \cdot \pi_{(h,w,r,q)} \quad (14)$$

$$\overline{r} = E[r] = \sum_{h=0}^{N}\sum_{w=0}^{N}\sum_{r=0}^{N}\sum_{q=0}^{K} r \cdot \pi_{(h,w,r,q)} \quad (15)$$

## IV. OPTIMIZATION METHOD

This section first presents the problem formulation and then the corresponding Markov decision process. Table 1 defines the variables to be used later.

### A. Optimization Problem Description

We define $X_{migr} = (\beta_h c_{hw} + \beta_w c_{wh})$, $X_{repair} = \mu_r c_r$ and $X_{host} = (\overline{h} c_h + \overline{w} c_w)$. Then Eq.(16) is the optimization function, where $\mu_r, \beta_h, \beta_w$ are variables. $\varpi_i$ is the weight.

$$\min X_{host}\varpi_1 + X_{repair}\varpi_2 + X_{migr}\varpi_3 + T_{resp}(1-\sum_{i=1}^{3}\varpi_i), \quad (16)$$

$$s.t. \quad \mu_r > \xi, \quad (17)$$

$$\beta_h, \beta_w \geq 0, \quad (18)$$

### B. NSGA-II -Based Method

The multi-objective evolutionary algorithm (MOEA) is a sophisticated technique for dealing with multi-objective optimization problems by mimicking the natural evolutionary process of a population. This technique searches solutions that approximate the Pareto frontier, a concept where no single solution can be deemed superior to another in the context of multiple objectives. A Pareto front solution means that is not dominated by any other solution within the feasible solution space. In the context of Practical Byzantine Fault Tolerance (PBFT), the Non-dominated Sorting Genetic Algorithm II (NSGA-II) [31] is applied based on MOEA. To understand this, we first introduce the fundamentals of the genetic algorithm (GA), which is the backbone of NSGA and other evolutionary algorithms.

### B.1 GA Basics

A Genetic Algorithm (GA) is a randomized search technique which applies the evolutionary ideas and is introduced by Hollan [32]. It is an optimization technique rooted in the principles of natural selection and genetics. It operates by:

• Initialization: An initial population of candidate solutions is created first, typically denoted as strings of binary digits (chromosomes).

• Evaluation: The fitness of each individual in the population based on a predefined objective function is evaluated.

• Selection: Individuals are selected for reproduction based on their fitness. Common selection methods include roulette wheel selection, tournament selection, and rank selection.

• Crossover: The genetic information of two parents is combined to produce offspring (children). This is akin to genetic recombination in biological organisms.

• Mutation: A candidate solution is randomly perturbed by changing the bits of an individual with a small predefined probability.

• Replacement: A new population is generated by replacing or augmenting the existing one with the newly created offspring.

In each generation, the values of a fitness function are computed for each individual. The algorithm stops when either a maximum number of generations are produced, or a satisfactory fitness level is reached for the population. Usually, the fitness function is the objective function of the problem to be handled. More details about GAs can be found in [32].

*B.2 Method Description*

In the designed scenario, we consider two goals: minimize maintenance cost and minimize the PBFT service response time. We need to determine the decision variables $\mu_r, \beta_h$ and $\beta_w$ to optimize the objective function in Eq. (16). These three variables are designed as the genes of population individuals. Algorithm 1 describes the NSGA-II-based method to seek the optimal solution, detailed in the following.

Initially, the target values $ObjV$ of the individuals within the initial population are computed. Subsequently, these individuals are stratified based on non-dominance, and their respective fitness values $FitnV$ are calculated. In this context, special consideration is given to the system's response time.

Then, elite individuals are selected as parents using the tournament method to undergo crossover and mutation to produce the next generation. The variables are encoded in real numbers, and we employ the Simulated Binary Crossover (SBX) operator and the polynomial Mutation (PM) operator. Eq. (19) represents the formula for the gene crossover operation.

$$y = 0.5*[(1-\psi)x_1 + (1+\psi)x_2] \qquad (19)$$

Here, $x_1$ and $x_2$ are the parent individuals. $\psi$ is a distribution factor, which determines the degree of similarity between the offspring and the parents, ranging from [-1, 1]. When $\psi$ is 0, the offspring is equal to the mean value of the parental genes. It means that the offspring is significantly different from one of the parents when it is close to 1 or -1.

The PM method operate as follows: a mutation parameter is randomly generated, and based on a polynomial distribution, a mutation value is derived. This mutation value is then calculated in conjunction with the gene value to effectuate individual variation.

Subsequently, the parent and offspring individuals are merged, and they are ranked according to both Pareto classification and crowding distance. We utilize a duplication operator to select and retain the elite individuals. The population undergoes iterative evolution following the aforementioned steps until the results converge, thereby obtaining the optimal solution.

**Algorithm 1** NSGA-II-based Method

**Input:** Genetic constraints, objective fuctions

**Output:** The optimal devision variables $\mu_r, \beta_h, \beta_w$

Initialize the population chromosomes, respectively

1: Compute the objective fuction value of the population $ObjV$
2: Non-dominant stratification of individuals
3: Compute the primary individual fitness $FitnV$
4: **While the evolutionary method is not terminated do**:
5:     Choose individuals to join evolution
6:     Conduct genetic crossover on c individuals
7:     Conduct genetic mutation on Conduct individuals
8:     Get the objtive fuction value of the evolved individual
9:     Reinsertion to create a new generation of populations
10: **End While**

## V. EXPERIMENT EVALUATION

This section first carries our numerical analysis to investigate how the variables of $N$, $\mu_r$, $\beta_h$, $\beta_w$ affect the PBFT service availability, the consensus processign delay and the dropping probability of the arriving transactions. Then we conduct simulations to valuate the optimization method.

*A. Numerical Analysis*

Table 1 gives the parameter default settings. Experiments are all carried out by using Python. We focus on performance by varying $N$ and varying one of the variables of $\mu_r, \beta_h, \beta_w$. Fig. 4 gives the results, detailed in the following subsections.

*A.1 Availability Analysis*

Fig.4.(a)-(c) show how the number of nodes ($N$) affects the PBFT service availability. We observe that an increase in the quantity of nodes enhances the PBFT service availability, but the extent of impact varies with different parameter settings.

- When $N$ is less than 15, thers is a notable enhancement in the PBFT service availability with the increase in $N$.
- When $N$ is greater than 15, the availability increases very slowly and gradually converges to a certain value. If $N$ keeps increasing, there will be no significant change in the availability either.
- The value of each of the parameter $\mu_r, \beta_h, \beta_w$ influences the upper limit of the availability. The black line in Fig.4(b) represents the result of setting $\mu_r$ =10, $\beta_h$ =0.2. With this setup, the availability gradually converges to 95% with increasing $N$.

*A.2 Performance Analysis*

Fig.4.(d)-(i) investigate the effect of $N$ on response time and dropping probability. As the number of nodes increases, bot the response time and dropping probability decrease. Note that the response time are of those transactions which are not dropped. The response time and the dropping probability are positively correlated.

- When $N$ is less than 15, the response time and the dropping probability decrease significantly as $N$ increases.
- When $N$ is greater than 15, the response time and the dropping probability decrease very slowly until they gradually approach their lower limit value. Subsequently, no significant change in the response time and the dropping probability is to be expected.
- The value of each of the parameter $\mu_r, \beta_h, \beta_w$ influences the lower limit of the response time. The black lines in Fig.4.(e) and (h) represent the results of setting $\mu_r=10$, $\beta_h=0.2$. As the number of nodes increases under this setup, the response time and the dropping probability gradually approach 1 and 1%, respectively..

In conjunction with the availability results, the response time and the availability are negatively correlated, as are the dropping probability and the availability.

*A.3 Effect of Parameter Settings*

Fig.4.(a), (d) and (g) illustrate the effect of different repair rate $\mu_r$ values on system perfromance. It can be concluded that for fixed node size $N$, the increasing repair rate leads to the results that the PBFT service availability increases, the response time and dropping probability decrease. The reason is explained as follows:

- As $\mu_r$ increases, once a node is sent to the RP from the HP, it is able to get repaired as soon as possible. Consequently, the number of failed nodes awaiting repair in the RP will decline and more repaired nodes will be sent to the WP. The WP is capable of maintaining an adequate number of standby nodes to ensure the continued support to the HP. When the number of active nodes in the HP does not satisfy the condition of greater than $3f+1$, there are always standby nodes in the WP that can be added to the HP.

Fig.4.(b), (e) and (h) illustrate the effect of different $\beta_w$ values on system perfromance. It can be concluded that for fixed node size $N$, the increasing $\beta_w$ leads to the results that the PBFT service availability increases, the response time and dropping probability decrease. The reason is explained as follows:

- As $\beta_w$ increases, once the number of active nodes in the HP is less than $3f+1$, standby noeds can be transferred from the WP to the HP with minimal delay. Therefore, the HP can maintain the continuity of the consensus over an extended period of time. In other words, there is always a sufficient number of active nodes in the HP to make consensus.

Fig.4.(c), (f) and (i) illustrate the effect of different $\beta_h$ values on system performance. It can be concluded that for fixed node size $N$, the increasing $\beta_h$ leads to the results that the PBFT service availability decreases, the response time and dropping probability increase. The reason is explained as follows:

- As $\beta_h$ increases, once the number of active nodes in the HP is greater than $3f+1$, the HP will proceed to transfer the nodes at a faster rate to the WP for the purpose of saving maintenance costs. This indicates that maintaining a high number of active nodes in the HP for a long period of time is not feasible. More nodes will remain in the WP on a more regular basis.

**B. Simulation Evaluation**

In this section, we first present the detailed settings of experiment parameters and our method parameters in Section B.1. Then, experiment results are given in Section B.2.

*B.1 Experiment Setting*

Based on the numerical analysis in Section A, we have assumed a failure rate of 0.5 and $c_h, c_w, c_r, c_{hw}, c_{wh}$ values are 5, 3, 2, 1, 1.5. Then we established the ranges for the decision variables $\mu_r, \beta_h, \beta_w$ as [1,10], [0.1,30], [3,15], respectively. Using these parameters, we proceed to solve the problem using a genetic algorithm. The initial popular-tion size is set to 50 individuals, and the maximum number of evolutionary generations is defined as 200.

All experiments are conducted with Geatpy on Ubuntu 20.04. The hardware configuration includes an Intel(R) Xeon(R) Gold 6230 CPU @ 2.10GHz, with 256G RAM and two NVIDIA RTX 4090 GPUs.

*B.2 Simulation Performance*

Our literature investigation indicates that there is no existing work considering our PBFT system. We utilize a genetic algorithm to tackle this problem and assess its performance in comparison to a random algorithm-based method. The random algorithm-based method operates as a basic heuristic method, randomly selecting a variable's value within a specified range and evaluating its corresponding target value.

By observing the trend variations depicted in Fig. 4, it becomes evident that the indicator undergoes significant changes prior to $N=15$ and gradually stabilizes by $N=20$. We conducted a comparative analysis of the proposed method and the random algorithm-based method at three specific instances: $N=12$, $N=15$, and $N=20$. As illustrated in Fig. 5, it is apparent that in all scenarios, our method performs better.

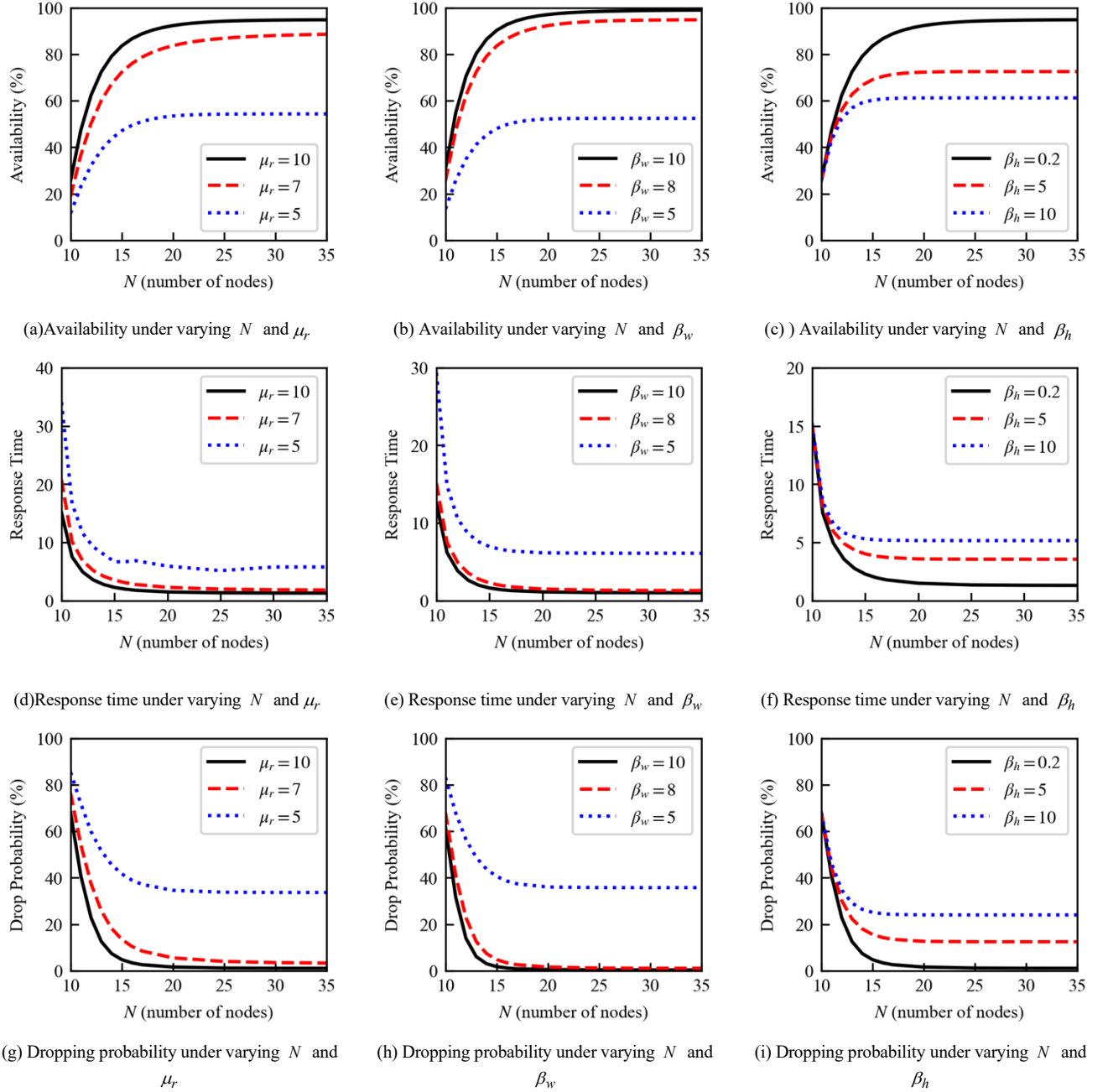

(a) Availability under varying $N$ and $\mu_r$
(b) Availability under varying $N$ and $\beta_w$
(c) Availability under varying $N$ and $\beta_h$
(d) Response time under varying $N$ and $\mu_r$
(e) Response time under varying $N$ and $\beta_w$
(f) Response time under varying $N$ and $\beta_h$
(g) Dropping probability under varying $N$ and $\mu_r$
(h) Dropping probability under varying $N$ and $\beta_w$
(i) Dropping probability under varying $N$ and $\beta_h$

Fig. 4 Performance over varying

To further demonstrate the superiority of the NSGA-II-based method, in the scenario where $N=15$, we fixed one of the decision variables while varying the other two. We then observed the resulting trends, as illustrated in Fig. 6. Specifically, Fig. 6 (a)、(b)、(c) fixes $\mu_r$、$\beta_h$、$\beta_w$ respectively, and change the other two variables. The **yellow** dot in each figure represents the obtained solution. By comparing these results, we observe that in all cases, the optimal solutions we obtain prioritize minimizing the system's response time, albeit at a cost. At this point, we get $\mu_r=9.9999$, $\beta_h=30$, $\beta_w=3$. If placing a higher emphasis on cost over response time is necessary, one can adjust the weighting coefficients for multiple objectives within the individual fitness evaluation of the population to better align with the optimization objectives.

Furthermore, we present the changes in two evaluation indicators of the NSGA-II-based method: HV (Hypervolume) and Spacing, as shown in Figure 7. The HV index measures the coverage supervolume of the evolutionary solution set relative to the Pareto front solution. A larger HV value indicates that the solution set is more convergent and closer to the Pareto front solution. On the other hand, the Spacing index represents the standard deviation of the distances between solutions within the set. A smaller Spacing value indicates a more uniform convergence and internal distribution of the solution set.

## VI. CONCLUSIONS AND FUTURE WORK

This paper aims for an effective PBFT consensus system under non-Byzantine errors, from the perspective of minizing the PBFT consensus processing time and the maintenance cost of PBFT consensus service. We first design a PBFT system, composed of three subsystems which collaborate to complete the consensus in time without causing much maintenance cost in the dynamic environment. We then develop a novel Markov-chain-based analytical model and the metric formulas for the quantitative analysis of the PBFT system performance and availability. The metrics include consensus processing time, PBFT service availability, the mean number of nodes in each subsystem. We also apply the deveopled metric formulas to set up the optimization problem for reducing both the PBFT service response time and the PBFT system maintenance cost. Finallly, We design a *N*ondominated *S*orting *G*enetic *A*lgorithm II (NSGA-II)-based method for the optimization problem. Numerical analysis and simulation experiments are conducted for evaluation.

This paper applies a monolithic model for capturing the dynamics of the proposed PBFT system. With the increasing number of consensus nodes, there exists the issue of state space explosion. One future research direction is to explore the hierarchical modeling technique for quantitatively analyzing the PBFT system. In addition, more optimization objectives will be studied.

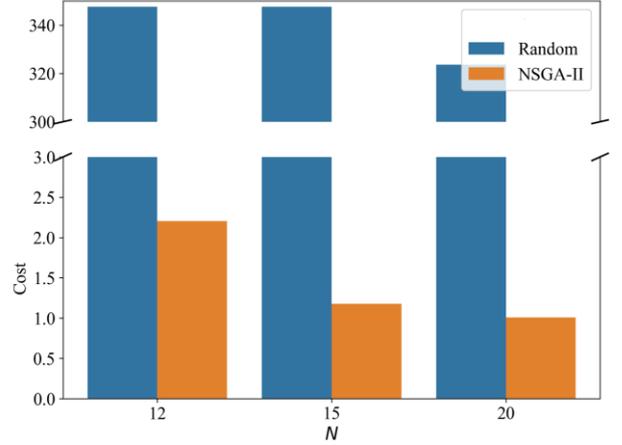

Fig.5 Comparison of the results

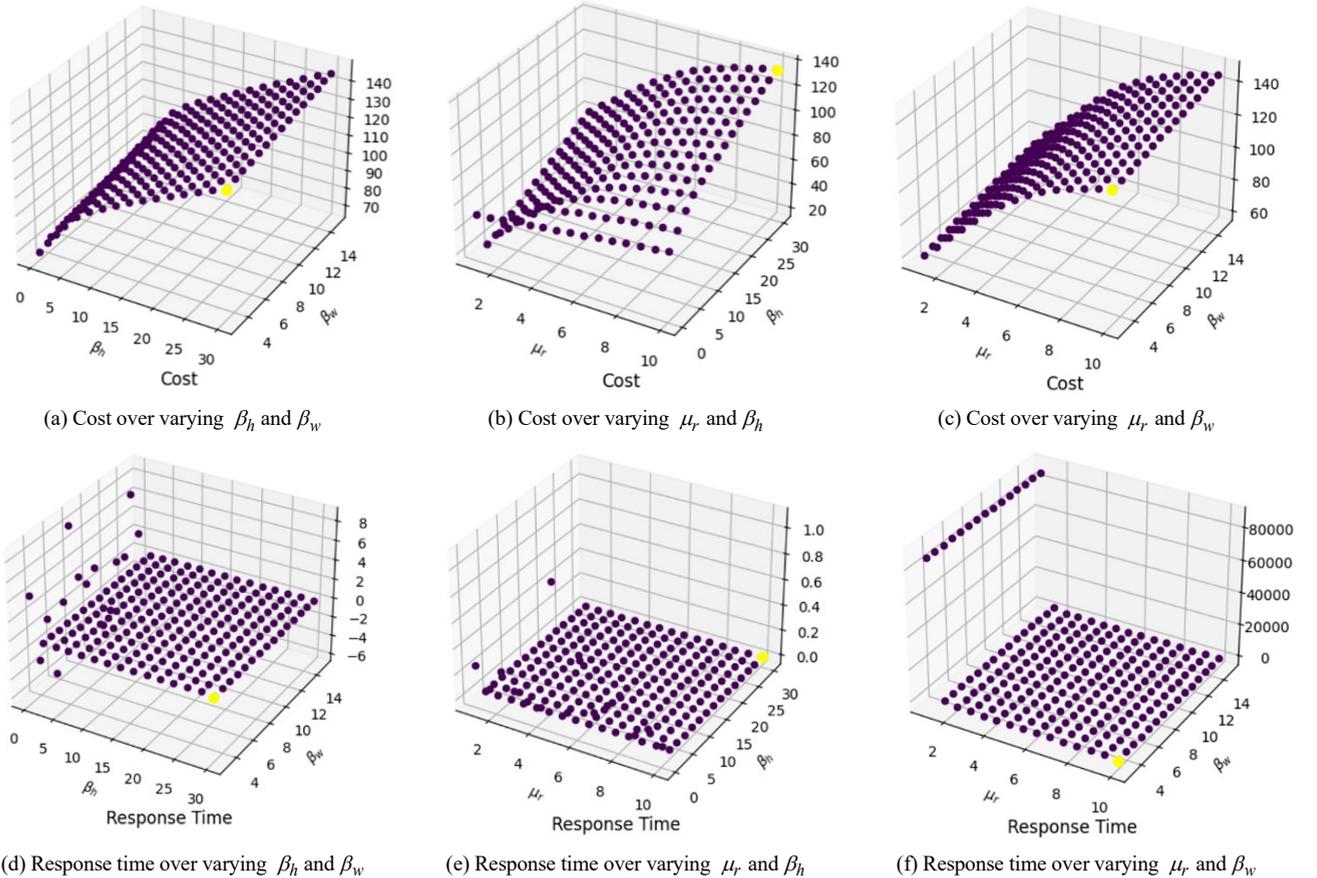

(a) Cost over varying $\beta_h$ and $\beta_w$  
(b) Cost over varying $\mu_r$ and $\beta_h$  
(c) Cost over varying $\mu_r$ and $\beta_w$  
(d) Response time over varying $\beta_h$ and $\beta_w$  
(e) Response time over varying $\mu_r$ and $\beta_h$  
(f) Response time over varying $\mu_r$ and $\beta_w$  

Fig.6 NSGA-II -based method performance

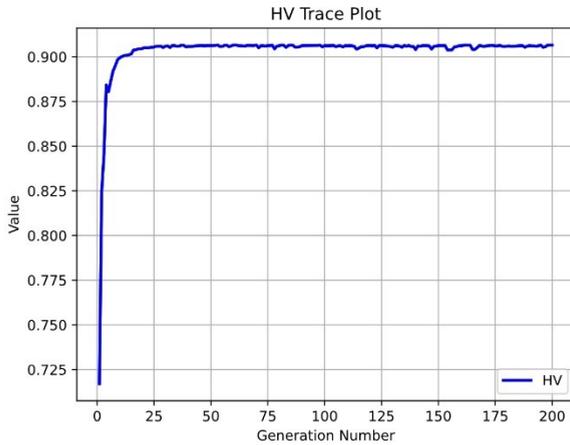

(a) HV Trace

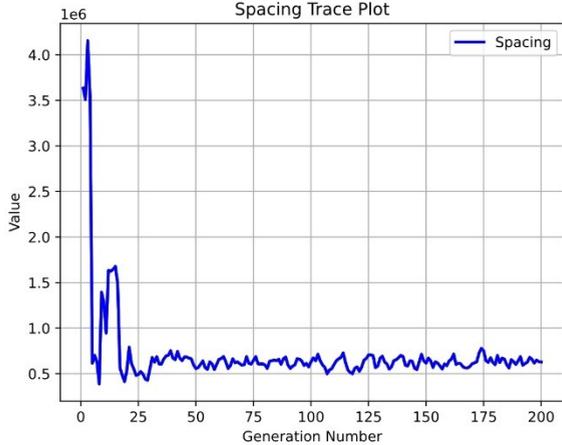

(b) Spacing Trace

Fig. 7 Evaluation Index of NSGA-II-based method

## Reference

[1] Hanjie Wu, Qian Yao, Zhenguang Liu, Butian Huang, Yuan Zhuang, Huayun Tang, Erwu Liu: Blockchain for finance: A survey. IET Blockchain 4(2): 101-123 (2024)

[2] Mikail Mohammed Salim, Laurence Tianruo Yang, Jong Hyuk Park: Privacy-preserving and scalable federated blockchain scheme for healthcare 4.0. Comput. Networks 247: 110472 (2024)

[3] Abeer Z. Al-Marridi, Amr Mohamed, Aiman Erbad: Optimized blockchain-based healthcare framework empowered by mixed multi-agent reinforcement learning. J. Netw. Comput. Appl. 224: 103834 (2024)

[4] Shaohua Cao, Zijun Zhan, Congcong Dai, Shu Chen, Weishan Zhang, Zhu Han: Delay-Aware and Energy-Efficient IoT Task Scheduling Algorithm With Double Blockchain Enabled in Cloud-Fog Collaborative Networks. IEEE Internet Things J. 11(2): 3003-3016 (2024)

[5] Chahrazed Benrebbouh, Houssem Mansouri, Sarra Cherbal, Al-Sakib Khan Pathan: Enhanced secure and efficient mutual authentication protocol in IoT-based energy internet using blockchain. Peer Peer Netw. Appl. 17(1): 68-88 (2024)

[6] Bhawana, Sushil Kumar, Rajkumar Singh Rathore, Upasana Dohare, Omprakash Kaiwartya, Jaime Lloret, Neeraj Kumar: BEET: Blockchain Enabled Energy Trading for E-Mobility Oriented Electric Vehicles. IEEE Trans. Mob. Comput. 23(4): 3018-3034 (2024)

[7] Hiren Dutta, Saurabh Nagesh, Jawahar Talluri, Parama Bhaumik: A Solution to Blockchain Smart Contract Based Parametric Transport and Logistics Insurance. IEEE Trans. Serv. Comput. 16(5): 3155-3167 (2023)

[8] Anthony Ugochukwu Nwosu, S. B. Goyal, Anand Singh Rajawat, Chaman Verma, Zoltán Illés: Enhancing Logistics With the Internet of Things: A Secured and Efficient Distribution and Storage Model Utilizing Blockchain Innovations and Interplanetary File System. IEEE Access 12: 4139-4152 (2024)

[9] Saqib Ali, Qianmu Li, Abdullah Yousafzai: Blockchain and federated learning-based intrusion detection approaches for edge-enabled industrial IoT networks: a survey. Ad Hoc Networks 152: 103320 (2024)

[10] Wei Chen, Wei Wu, Zhiyuan Ouyang, Yelin Fu, Ming Li, George Q. Huang: Event-based data authenticity analytics for IoT and blockchain-enabled ESG disclosure. Comput. Ind. Eng. 190: 109992 (2024).

[11] Xin Wang, Sisi Duan, James R. Clavin, Haibin Zhang: BFT in Blockchains: From Protocols to Use Cases. ACM Comput. Surv. 54(10s): 209:1-209:37 (2022)

[12] Gengrui Zhang, Fei Pan, Yunhao Mao, Sofia Tijanic, Michael Dang'ana, Shashank Motepalli, Shiquan Zhang, Hans-Arno Jacobsen: Reaching Consensus in the Byzantine Empire: A Comprehensive Review of BFT Consensus Algorithms. ACM Comput. Surv. 56(5): 134:1-134:41 (2024)

[13] Jiamou Qi, Yepeng Guan: Practical Byzantine fault tolerance consensus based on comprehensive reputation. Peer Peer Netw. Appl. 16(1): 420-430 (2023)

[14] Haoxiang Luo, Xiangyue Yang, Hongfang Yu, Gang Sun, Bo Lei, Mohsen Guizani: Performance Analysis and Comparison of Nonideal Wireless PBFT and RAFT Consensus Networks in 6G Communications. IEEE Internet Things J. 11(6): 9752-9765 (2024)

[15] Lili Jiang, Xiaolin Chang, Yuhang Liu, Jelena V. Misic, Vojislav B. Misic: Performance analysis of Hyperledger Fabric platform: A hierarchical model approach. Peer-to-Peer Netw. Appl. 13(3): 1014-1025 (2020).

[16] Jelena V. Misic, Vojislav B. Misic, Xiaolin Chang:Design of Proof-of-Stake PBFT Algorithm for IoT Environments. IEEE Trans. Veh. Technol. 72(2): 2497-2510 (2023)

[17] Yan-Xia Chang, Qing Wang, Quan-Lin Li, Yaqian Ma, Chi Zhang: Performance and Reliability Analysis for PBFT-Based Blockchain Systems With Repairable Voting Nodes. IEEE Trans. Netw. Serv. Manag. 21(4): 4039-4060 (2024)

[18] Jing Bai, Xiaolin Chang, Fumio Machida, Lili Jiang, Zhen Han, Kishor S. Trivedi: Impact of Service Function Aging on the Dependability for MEC Service Function Chain. IEEE Trans. Dependable Secur. Comput. 20(4): 2811-2824 (2023)

[19] Xiaolin Chang, Xiuming Mi, Jogesh K. Muppala: Performance evaluation of artificial intelligence algorithms for virtual network embedding. Eng. Appl. Artif. Intell. 26(10): 2540-2550 (2013).

[20] Caixiang Fan, Sara Ghaemi, Hamzeh Khazaei, Petr Musílek: Performance Evaluation of Blockchain Systems: A Systematic Survey. IEEE Access 8: 126927-126950 (2020)

[21] Mohammadreza Rasolroveicy, Wejdene Haouari, Marios Fokaefs: BlockCompass: A Benchmarking Platform for Blockchain Performance. IEEE Trans. Computers 73(8): 2111-2122 (2024)

[22] Jingyu Hao: A Blockchain-based Performance Monitoring Scheme for Cloud Applications. CNCIT 2023: 201-207

[23] Jiashuo Zhang, Jianbo Gao, Zhenhao Wu, Wentian Yan, Qize Wu, Qingshan Li, Zhong Chen: Performance Analysis of the Libra Blockchain: An Experimental Study. CoRR abs/1912.05241 (2019)

[24] Suporn Pongnumkul, Chaiyaphum Siripanpornchana, Suttipong Thajchayapong: Performance Analysis of Private Blockchain Platforms in Varying Workloads. ICCCN 2017: 1-6

[25] Harish Sukhwani, José M. Martínez , Xiaolin Chang, Kishor S. Trivedi, Andy J. Rindos: Performance Modeling of PBFT Consensus Process for Permissioned Blockchain Network (Hyperledger Fabric). SRDS 2017: 253-255

[26] Fan-Qi Ma, Quan-Lin Li, Yi-Han Liu, Yan-Xia Chang: Stochastic performance modeling for practical byzantine fault tolerance consensus in the blockchain. Peer-to-Peer Netw. Appl. 15(6): 2516-2528 (2022)

[27] Thomas Lorünser, Benjamin Rainer, Florian Wohner: Towards a Performance Model for Byzantine Fault Tolerant Services. CLOSER 2022: 178-189

[28] Jelena V. Misic, Vojislav B. Misic, Xiaolin Chang, Haytham Qushtom: Adapting PBFT for Use With Blockchain-Enabled IoT Systems. IEEE Trans. Veh. Technol. 70(1): 33-48 (2021)

[29] Haytham Qushtom, Jelena V. Misic, Vojislav B. Misic, Xiaolin Chang: A Two-Stage PBFT Architecture With Trust and Reward Incentive Mechanism. IEEE Internet Things J. 10(13): 11440-11452 (2023)

[30] Marco Marcozzi, Leonardo Mostarda: Analytical model for performability evaluation of Practical Byzantine Fault-Tolerant systems. Expert Syst. Appl. 238(Part A): 121838 (2024)

[31] Kalyanmoy Deb, Samir Agrawal, Amrit Pratap, T. Meyarivan: A fast and elitist multiobjective genetic algorithm: NSGA-II. IEEE Trans. Evol. Comput. 6(2): 182-197 (2002).

[32] John H. Holland. Adaptation in Natural and Artificial Systems. University of Michigan Press, AnnArbor, MI, 1975.

[33] Weijie Zheng, Benjamin Doerr: Runtime Analysis for the NSGA-II: Proving, Quantifying, and Explaining the Inefficiency for Many Objectives. IEEE Trans. Evol. Comput. 28(5): 1442 - 1454 (2024)